# A New Chaos and Permutation Based Algorithm for Image and Video Encryption

Chinmaya Patnayak
Birla Institute of Technology & Science, Pilani - Hyderabad
Hyderabad, India
chinmayapatnayak@gmail.com

Pradipta Roy
Integrated Test Range,
Defense Research & Development Organization (DRDO)
Balasore, India
pradiptar@yahoo.com

Bibekanand Patnaik
Integrated Test Range,
Defense Research & Development Organization (DRDO)
Balasore, India
bpatnaik.drdo@gmail.com

*Abstract*— **Images and video sequences carry large volumes of highly correlated and redundant data. Applications like military and telecommunication require encryption methods to protect the data from unwanted access. This requirement in most cases needs to be realized in real-time. In this paper, we propose a fast new Fiestal-structured approach for image and video encryption based on a chaotic random sequence generator and a Permutation-Inverse Permutation (PIP) pixel transform. This approach utilizes mathematical functions and transforms with low complexity. The algorithm at the same time, ensures no drastic pay off in terms of encryption quality. This renders the algorithm with promising scope for real time applications and easy hardware implementation. MATLAB simulation of the algorithm establishes its high quality of encryption in terms of elevated entropy values and negligible correlation of the encrypted data with the original. Simulation results also show high sensitivity to slight variation in keys ensuring high security.**

## I. INTRODUCTION

Exchange of information through wired and wireless channels and networks has bloomed and magnified over the years. The progress in wide ended means of communication, including the Internet and satellite, stand out as exemplary. With such a trend in its growth, security and confidentiality of essential information to be transmitted or stored, has emerged as a prominent concern. Applications including military imaging and video streaming, medical imaging, video conferencing, and telecommunication find themselves at the heart of attacks from hackers and eavesdroppers.

Encryption methods are tools for transforming the correlated pieces of information into an unrecognizable format. A lot of related research has been undertaken over the past years. The classification of existing encryption methods on the basis of their effect on a plain-text, as earlier discussed in [1], can be done as : (1) Value Transformation, [2] (2) Position Permutation, [3] or (3) a combination of both [4]. In context of video/image encryption, the value transformation method transforms the pixel value from the original. Whereas, the position permutation method scrambles the positions of every pixel.

A few stand-out encryption algorithms include the Triple-Data Encryption Standard (3DES) and Advanced Encryption Standards (AES). Owing to the high quality of encryption they offer, they are now used as standards. But the high complexity associated with them and real time constraints, makes them unsuitable for image and video encryption which carry large volumes of data. Also, digital images and videos tend to maintain a high level of correlation between adjacent pixels and/or frames, which results in failure of many other well known encryption techniques.

Chaos, due to its noise like properties and deterministic randomness has found prominent significance in the field of cryptography. Bourbakis and Alexopaulus have introduced SCAN language to encrypt and compress an image simultaneously [2]. In 1999, Yen and Guo proposed the BRIE scheme which utilized bit-wise gray manipulations in addition to usage of chaotic logistic map [3]. J.C. Yen et al, in 2000, demonstrated the usage of XOR with chaotic keys in their research [5]. Guodong Ye, proposed an algorithm for image encryption by scrambling of pixel bits based on a chaos map [6]. Chaos has also successfully been extended to video surveillance domain.

This paper introduces a enhanced and faster new scheme for multimedia encryption. This Fiestal-structured scheme is a novel amalgamation of a chaotic random number generator, based on a 1-D logistic map and a new pixel-bit level permutation transformation [7]. The chaotic sequences provide the necessary confusion into the image/video. Diffusion, on the other hand is introduced, with a permutation & inverse permutation (PIP) technique that scrambles pixel bits with respect to its position and association with its transformed neighbors. This PIP mapping also adds an additional layer to the existing confusion.

The paper is divided into five parts. In Section II, the new chaotic encryption scheme is proposed. The algorithm and its features are discussed in section III. Computer

`

simulation and security analysis will be shown in Section IV. In Section V, a conclusion will be reached.

## II. ALGORITHM

This section introduces the algorithm of the new chaos and permutation encryption scheme.

Let f be the associated pixel value corresponding to its position given by (x,y,z), where $0 \leq f \leq 255$, $0 \leq x \leq L$, and $0 \leq y \leq B$, $0 \leq z \leq 3$ for a frame of image in RGB format of the dimensions LxB.

1) The initial key $\chi(0)$ and the parameter $\mu$ is obtained from the central key for the image, G(f(x)).

2) A pixel block of 8 consecutive pixels, $g_0(f(x))$, each pixel value represented by 8 bits, is selected. A bit-wise position scrambling is done according to a position permutation table. This results in a different 8 pixel block, say $g_0^T(f(x))$.

3) For each of the pixels from the block $g_0^T(f(x))$, a key, $\chi(n)$ is generated.

4) A bit-wise XOR is performed on each pixel bit with the 8 MSBs of the $\chi(n)$.

5) The bit-wise XOR process of all the pixels from the pixel block $g_0^T(f(x))$ with their corresponding keys transforms into a new block, say $g_0'^T(f(x))$.

6) Yet another bit-wise position scrambling (similar to step (2) with an inverse permutation table is performed to restore the block into $g_0'(f(x))$.

7) The next pixel blocks $g_1'(f(x))$, $g_2'(f(x))$, .. $g_n'(f(x))$, are selected and the operations as detailed in steps 1-6 are performed, to generate the encrypted image G'(f(x)).

The decryption process just requires the encrypted image G'(f(x)) to be put in as the input. When processed with the correct key, the algorithm generates the decrypted image as the output.

The block diagram for the algorithm is shown in Fig. 1.

## III. ALGORITHM ANALYSIS

A Chaotic system, as is known [7], is a deterministic system with random behavior. Their high sensitivity to initial conditions and stable yet non periodic or quasi periodic properties, makes it a mathematically and statistically suitable tool for generating random numbers.

A well known system which shows chaotic behavior is the 1-D Logistic Map, with parameter $\mu$, which maps from the unit interval [0,1] into [0,1] :

$$f_\mu(\chi) = \chi\mu(1-\chi) \quad (1)$$

where, $0 \leq \mu \leq 4$, and the evolution of the function is given by :

$$f_\mu(\chi(n+1)) = \chi(n)\mu(1-\chi(n)) \quad (2)$$

Equation (2) is used as a generating function for the keys, $\chi(n)$, with the system initialized with the values for $\mu$ and $\chi(0)$.

Even with the chaotic behavior taking care of delivering the necessary confusion into an image/video frame, there still exists a considerable correlation amongst neighboring pixels. This in most cases resulted in texture zones and patterns, the effect assuming a more prominent form with changing frames. Such an effect is shown in Fig. 3 (c)-(d), where a test image is pixel-by-pixel processed with XOR function with the keys generated from the 1-D chaos logistic map. Texture zones can be easily identified in the encrypted image.

The 1-D Logistic Mapping function also showed some bias in generation of random number when tested for a set of initial test values. One such histogram for $\chi(0) = 0.22101986$, for the next 100,000 key values is represented in Fig. 2.

A second layer of confusion is ensured by implementing a Permutation and Inverse Permutation technique. Wherein pixel bits are scrambled within a pixel block, that also renders diffusion properties into the image/video frame. So, a particular pixel, if needed to be decrypted, cannot be done, until the mapping with its adjacent pixel is known.

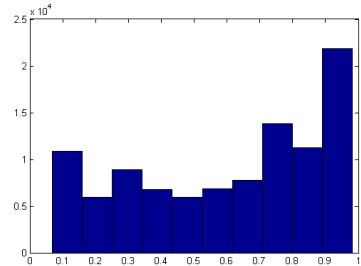

Fig. 2. Histogram depicting distribution of $\chi(n)$ for $\chi(0) = 0.22101986$ for 100,000 evolutions

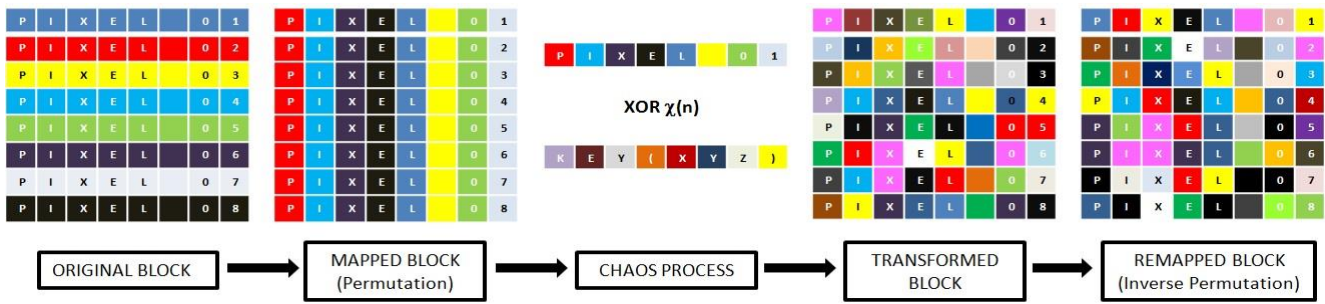

Fig. 1. The Chaotic Permutation Inverse Permutation Algorithm.

Thus, the algorithm, owing to the PIP and XOR operations, behaves as a 'Fiestal cryptosystem' wherein one station can act as both the encryption and decryption centre. This eliminates the need for development of a separate algorithm for encryption and decryption.

## IV. SIMULATION AND SECURITY ANALYSIS

The proposed algorithm is implemented in MATLAB. Various standard and non-standard images and video sequences were tested with this algorithm for encryption and decryption. The test results were evaluated for efficiency and quality based on various measures.

In qualitative analysis, preliminary manual observations were made on the encrypted image and video sequences which couldn't lead to recognition of any visible remnant patterns or decipherable spots.

For quantitative evaluation of the same, the following parameters are evaluated :

### A. Histogram Analysis

The histogram of an image or a video frame depicts the distribution of pixel content throughout its dimension. The histogram plots the number of pixels with a particular value against the value itself.

A sample grayscale test image is used for analyzing the behavior of the histogram, and the results are represented in (3). We observed that the histogram for the encrypted image is significantly different from the original image under test. The fairly uniform and normalized nature of the histogram of the encrypted image implies the algorithm's superior resistance to known statistical attacks as compared to the algorithm proposed by Guodong Ye [6] which attained non uniformly distributed peaks.

### B. Image Entropy

Entropy is the statistical measure of randomness of information. The entropy of an image is a function of the histogram count and is given as :

$$S(I) = -\sum_{i=0}^{255} p(x_i) \log_2(p(x_i)) \qquad (3)$$

where $p(x_i)$ is the sum of pixel counts for a particular grayscale value.

The results of the entropy analysis, as performed on five sample images, is tabulated in Table I. The initializing parameters were randomly set to : μ = 3.934 and χ(0) = 0.5250. From the table we observed that the entropy values have increased significantly close to 8, which represents very high level of randomness imparted to the encrypted image.

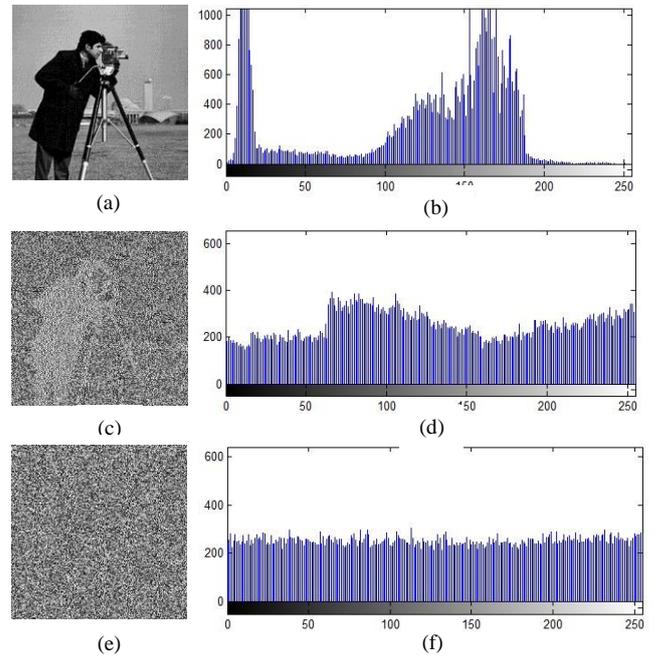

Fig. 3. Simulation results of the proposed algorithm : (a) the original image (b) the histogram for the original image (c) the encrypted image (d) the histogram for the original image

TABLE I. ENTROPY

| Image Title | Entropy of original image | Entropy of encrypted image |
|---|---|---|
| Lena | 7.4254 | 7.9973 |
| Peppers | 7.5485 | 7.9974 |
| Mandril | 7.2382 | 7.9968 |
| Livingroom | 7.3980 | 7.9970 |
| Cameraman | 7.0272 | 7.9967 |

### C. Correlation Coefficient

The correlation between adjacent pieces of information makes a image or video frame different from any other piece of plain text. The 2-D correlation coefficient were found out for the same five test images and initializing parameters as is used for computing entropy and is tabulated in Table II. We found that the correlation between the original image and the encrypted image was significantly destroyed as the 2-D correlation coefficients tends to zero. The correlation coefficients also are found to be smaller than the method proposed by Guodong Ye [6].

TABLE II. 2-D CORRELATION COEFFICIENTS

| Image Title | 2-D correlation coefficient (original image) | 2-D correlation coefficient (encrypted image) |
|---|---|---|
| Lena | 0.9913 | 0.0086 |
| Peppers | 0.9960 | 0.0001 |

| Image Title | 2-D correlation coefficient (original image) | 2-D correlation coefficient (encrypted image) |
|---|---|---|
| Mandril | 0.9373 | 0.0036 |
| Livingroom | 0.9770 | -0.0099 |
| Cameraman | 0.9924 | 0.0136 |

*D. Key Sensitivity*

The algorithm is tested for key sensitivity by encrypting an image by keys changed by a very small value. Two sample images are tested with $\chi(0) = 0.919666837573$ and $\chi'(0) = 0.919666837572$ and the correlation between the two encrypted images was found out and is tabulated in Table III. We observed that a slight change in the key resulted in a huge difference in the encryption pattern as the 2-D correlation coefficient again tends to zero.

TABLE III.  2-D CORRELATION COEFFICIENTS

| Image Title | 2-D correlation coefficient ($\chi(0)$ & $\chi'(0)$) |
|---|---|
| Lena | -0.0022 |
| Peppers | -0.0051 |

Similar tests were conducted for RGB test images and video sequences against all the parameters as described above. The results of encryption and decryption processes for the test image 'Lena' is shown in Fig. 4. The histograms of the original and encrypted images are compared individually for red, green and blue image frames, to demonstrate the quality of encryption. The original and encrypted frame from a video sequence, 'xylophone' is shown in Fig. 5.

## V. CONCLUSION

In this paper, we have introduced a new modified version of the existing methods on chaotic video and image encryption using 1-D logistic maps. The special bitwise Permutation & Inverse Permutation (PIP) mapping technique which has been incorporated in addition to XOR manipulations, ensures the low complexity of the algorithm without any significant loss in encryption quality. The analysis and exhaustive evaluation of the algorithm with respect to well established measures was undertaken. The results as collected and presented in the previous section establishes the method as an efficient and high security cryptosystem.

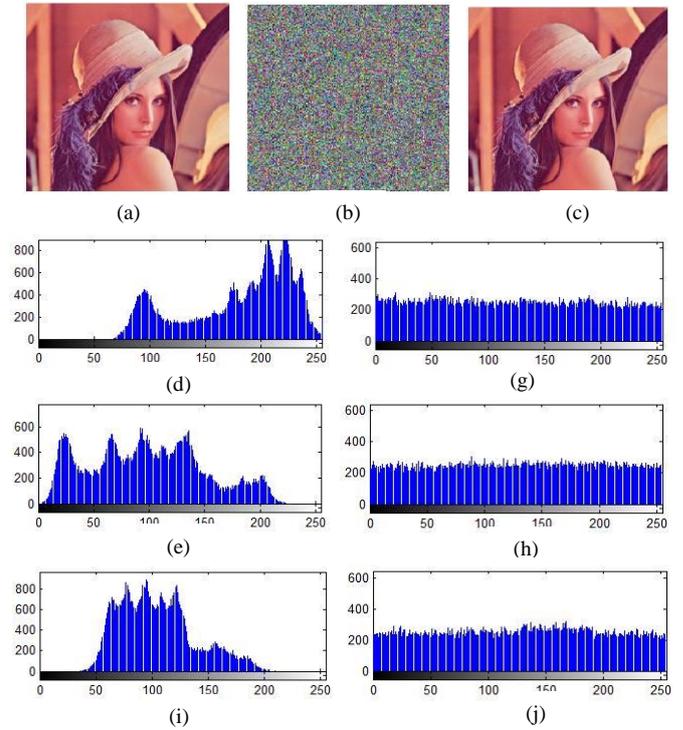

Fig. 4. Simulation results of the proposed algorithm on an RGB image : (a) the original image (b) the encrypted image (c) the decrypted image (d)-(f) the histograms for the original image in the order red, green and blue frames (g)-(i) the histograms for the encrypted image in the order red, green and blue frames

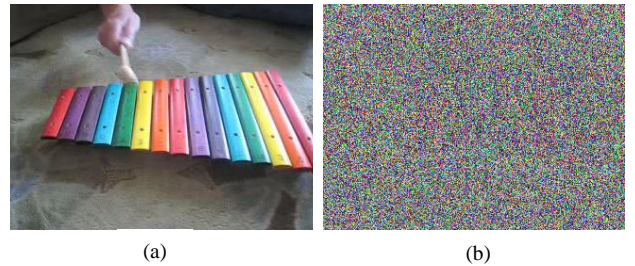

Fig. 5. Simulation results of the proposed algorithm on test video : (a) the original video frame (b) the encrypted video frame